\documentclass[prx,superscriptaddress,twocolumn]{revtex4-1}

\usepackage[english]{babel}  
\usepackage[]{graphics,graphicx,epsfig}
\usepackage{amsmath}
\usepackage{dcolumn,txfonts}

\newcommand{\trace}{\mbox{Tr}}

\begin{document}

\title{The relation between nonlocality and contextuality for a biphoton}
\author{Akihito Soeda}
\affiliation{Centre for Quantum Technologies, National University of Singapore, 3 Science Drive 2, 117543 Singapore, Singapore}

\author{Pawe{\l} Kurzy\'nski}
\affiliation{Centre for Quantum Technologies, National University of Singapore, 3 Science Drive 2, 117543 Singapore, Singapore}
\affiliation{Faculty of Physics, Adam Mickiewicz University, Umultowska 85, 61-614 Pozna\'{n}, Poland} 

\author{Ravishankar Ramanathan}
\affiliation{Centre for Quantum Technologies, National University of Singapore, 3 Science Drive 2, 117543 Singapore, Singapore} 

\author{Kavan Modi}
\affiliation{Department of Physics, University of Oxford, Clarendon Laboratory, Oxford United Kingdom}
\affiliation{Centre for Quantum Technologies, National University of Singapore, 3 Science Drive 2, 117543 Singapore, Singapore}

\author{Dagomir Kaszlikowski}
\email{phykd@nus.edu.sg}
\affiliation{Centre for Quantum Technologies, National University of Singapore, 3 Science Drive 2, 117543 Singapore, Singapore}
\affiliation{Department of Physics, National University of Singapore, 2 Science Drive 3, 117542 Singapore, Singapore}

\begin{abstract}
We investigate the set of qutrit states in terms of symmetric states of two qubits that violate the minimal contextual inequality, namely the Klyachko-Can-Binicoglu-Shumovsky (KCBS) inequality. The physical system that provides a natural framework for this problem is a biphoton which consists of two photons in the same spatio-temporal mode and whose effective polarization behaves as a three-level quantum system. The relationship between the KCBS contextual inequality and the Clauser-Horne-Shimony-Holt (CHSH) inequality is investigated. We find that every biphotonic state that is contextual with respect to KCBS is nonlocal as per the CHSH test when the two photons are apart, but the converse is not true. 
\end{abstract}

\maketitle

\section{Introduction}

The concept of contextuality states that the outcomes of measurement may depend on what measurements are performed alongside. Perceived in this light, nonlocality is a special instance of contextuality, where simultaneous measurements are facilitated by spatial separation. The smallest system exhibiting nonlocality consists of two qubits on which one may perform the well-known Clauser-Horne-Shimony-Holt (CHSH) tests~\cite{CHSH}, these being the simplest nonlocality tests for this system. Analogously, the smallest contextual system is a qutrit with the simplest test of contextuality being the Klyachko-Can-Binicoglu-Shumovsky (KCBS) inequality~\cite{KCBS}. Since a single qutrit can be represented as a system of two qubits, it is natural to investigate the relationship between the set of quantum states that exhibit nonlocality and the set that display contextual behavior.

Consider a single qutrit composed of two qubits in a symmetric state in two experimental scenarios. In the first scenario, access to individual qubits is forbidden and the only available transformations are restricted to joint $SO(3)$ operations on the Block vectors of both qubits to test the contextuality of this system. This system is often represented in the literature by a spin-1 particle composed of two spin-1/2 particles. In this article we consider biphotons--- two photons in the same spatio-temporal mode--- which can be readily created~\cite{SO31, SO32, SO33, SO34, Aichele, Mikami}. Note, that mathematical description of biphoton is the same as the one of spin-1 composed of two spin-1/2 particles. The magnetization of spin-1/2 represented by spin operators corresponds to the polarization of a single photon and magnetization of spin-1 corresponds to the polarization of biphoton. Since photons are indistinguishable bosons, their joint polarization state is restricted to the symmetric three-dimensional Hilbert space. Joint $SO(3)$ operations on both photons can be implemented by linear optics whereas more complicated general $SU(3)$ transformations require interactions between photons that need inefficient non-linear phenomena~\cite{SO31, SO32, SO33, SO34, Aichele, Mikami}. 

In the second scenario the qubits are spatially separated and locally accessible, therefore amenable to Bell inequality tests. Note that this scenario is related to the previous one via the analogy to the famous Bohm formulation of the EPR paradox in which a single spin-1 particle splits into two spin-1/2 particles. Moreover, it naturally fits the biphotonic system --- one may simply split a single biphoton into two photons using a beam splitter without affecting the polarization state  (and postselect on these events when two photons actually split). One may then ask if a quantum state that violates a CHSH inequality in the latter scenario also violates the KCBS inequality in the former scenario, and vice versa. We find that all local states are non-contextual while intriguingly {\it there exist quantum states that are non-local yet non-contextual}. 

\section{Biphoton states}

We begin by finding the most general state of a biphoton system. Such a system is invariant under particle swap and symmetric in phase as well. The general state of two qubits (polarization of two photons) can be written as 
\begin{gather}
\rho=\frac{1}{4}\left(\mathbb{I}\otimes\mathbb{I} + \vec{a}\cdot\vec{\sigma}\otimes \mathbb{I} + \mathbb{I}\otimes \vec{b}\cdot\vec{\sigma} +\sum_{k,l=x,y,z} \hat{T}_{kl} \sigma_k \otimes \sigma_l \right),
\end{gather}
where $\vec{a}$ and $\vec{b}$ are local Bloch vectors and $\hat{T}_{kl}$ is the correlation tensor. The full description of the state is given by 15 real parameters in total. This state can be symmetrized by setting $\vec{a}=\vec{b}$ and $\hat{T}_{kl}=\hat{T}_{lk}$. Further we require that this state also has to be orthogonal to the antisymmetric subspace which in this case is one-dimensional and is represented by the singlet state 
\begin{gather}
|\psi_-\rangle=\frac{1}{\sqrt{2}}(|01\rangle-|10\rangle). 
\end{gather}
This orthogonality criterion gives 
\begin{gather}
\langle \psi_- |\rho_s |\psi_-\rangle = \frac{1}{4} \left( 1 - \hat{T}_{xx} - \hat{T}_{yy} - \hat{T}_{zz} \right) = 0, 
\end{gather}
which implies $\trace[\hat{T}]=1$. The resulting generic biphoton state has the form 
\begin{gather}
\rho_s=\frac{1}{4}\left(\mathbb{I} \otimes \mathbb{I} + \vec{a}\cdot(\vec{\sigma}\otimes \mathbb{I} + \mathbb{I} \otimes \vec{\sigma}) +\sum_{k,l} (\vec{k}^\intercal \cdot \hat{T} \cdot \vec{l}) \sigma_k \otimes \sigma_l \right). 
\end{gather}
This state is fully described by eight parameters: three corresponding to local Bloch vector and five corresponding to the symmetric correlation tensor $\hat{T}$~\cite{Usha}. Three out of five parameters of $\hat{T}$ correspond to off-diagonal terms. 

We are interested in the nonclassicality of this state as determined by CHSH and KCBS inequalities. First note that the optimal CHSH operator has the form
\begin{gather}\label{BCHSH}
B_{m,m_{\perp}}=\sqrt{2}\left(\vec{m}\cdot\vec{\sigma}\otimes \vec{m}\cdot\vec{\sigma}+\vec{m_{\perp}}\cdot\vec{\sigma}\otimes \vec{m_{\perp}}\cdot\vec{\sigma}\right),
\end{gather}
where $\vec{m}\cdot\vec{m_{\perp}}=0$. Next, when the polarization of biphoton is singularly measured, due to the rules of addition of angular momenta, the corresponding spin-1 operator has the following form in terms of the two corresponding spin-1/2 operators
\begin{gather}\label{3obs}
S_m=\frac{1}{2}\left(\vec{m}\cdot\vec{\sigma} \otimes \mathbb{I} + \mathbb{I} \otimes \vec{m}\cdot\vec{\sigma}\right).
\end{gather}
Therefore, the local Bloch vector $\vec{a}$ denotes the magnetization of the spin-1 system (biphotonic polarization) composed of two spin-1/2 particles (polarization of two photons). On the other hand
\begin{gather}\label{square}
S_m^2=\frac{1}{2}(\mathbb{I} \otimes \mathbb{I} +\vec{m}\cdot\vec{\sigma} \otimes \vec{m}\cdot\vec{\sigma}).
\end{gather}
It follows, that operators involved in the CHSH test are related to $S^2_m$ not to the first power of magnetization (polarization) which is described by local coherence vector $\vec{a}$. As we shall see shortly, the operators that we have to measure to determine the value of KCBS inequality have the form $\langle \sum_m S^2_m\rangle$. Therefore, neither CHSH or KCBS inequalities depend on the value of local coherence and we are free to set $\vec{a}=\vec{0}$, allowing for a maximal range for the eigenvalues of $\hat{T}$.

We can describe relevant states of a biphoton solely by the symmetric correlation tensor $\hat{T}$. Since it is symmetric and real, it is therefore Hermitian and can be diagonalized. In particular, we show that all correlation tensors whose eigenvalues obey $\lambda_1+\lambda_2+\lambda_3=1$ and $-1\leq \lambda_i \leq 1$, correspond to proper quantum states, i.e., positive semidefinite density matrices. To see this, note that the diagonal form of such state can be always written as 
\begin{gather}
\rho_s=\frac{1}{4} \left(\mathbb{I} \otimes \mathbb{I} + \lambda_1 \sigma_x \otimes \sigma_x + \lambda_2 \sigma_y \otimes \sigma_y + \lambda_3 \sigma_z \otimes \sigma_z\right).
\end{gather}
We can compute the eigenvalues of $\rho_s$, which gives four positivity criteria 
\begin{align}
& 1 + \lambda_1 + \lambda_2 - \lambda_3 \geq 0,\quad 
1 - \lambda_1 + \lambda_2 + \lambda_3 \geq 0,\\
& 1 + \lambda_1 - \lambda_2 + \lambda_3 \geq 0,\quad
1 - \lambda_1 - \lambda_2 - \lambda_3 \geq 0. 
\end{align}
Taking into account $\trace[\hat{T}]=1$ we get $\lambda_i \leq 1$. There always exists a valid quantum state corresponding to $\hat{T}$. Note, that the eigenvalues of $\hat{T}$ can be always ordered $1 \geq \lambda_1 \geq \lambda_2 \geq \lambda_3 \geq -1$. Due to $\trace[\hat{T}]=1$ only $\lambda_3$ can be negative, since negativity of $\lambda_2$ would imply $\lambda_1 > 1$. Moreover, $|\lambda_2|\geq |\lambda_3|$ due to the same reason.

\section{Biphoton manipulation} 

The standard biphoton generation procedure exploits spontaneous parametric down-conversion (SPDC) of type I or II. These processes can generate biphotons in a state in which polarization of both photons is the same (separable state) or opposite (maximally entangled state)~\cite{SO33}. An example of such states are either $|H,H\rangle$ and $|V,V\rangle$, or $\frac{1}{\sqrt{2}}(|H,V\rangle + |V,H\rangle)$, where $H$ and $V$ denote horizontal and vertical polarization, respectively. Following Refs.~\cite{SO33, SO34} we only consider operations that can be implemented via linear optics. These operations address the polarization degree of freedom of both photons in the same way and are mathematically described by $SO(3)$ group. It is clear that $SO(3)$ transformations cannot change entanglement between two photons, therefore the set of states that can be obtained from the original one generated via SPDC I or II is limited to the ones with the same amount of entanglement. Note, that due to the problems with operational meaning of entanglement in the systems of indistinguishable particles some authors refer to different types of biphoton states using the notion of {\it polarization degree}~\cite{SO34}.

\section{Contextuality for qutrits}

We focus on the state-dependent contextual inequalities of the Klyachko-Can-Binoicoglu-Shumovsky (KCBS) type~\cite{KCBS} which are of the form
\begin{gather} 
\label{KCBS1}
\sum_{m=1}^5 \langle P_m \rangle \leq 2.
\end{gather}
Here, the measurements $P_m$ (with measurement outcomes $0, 1$) are mutually compatible in a cyclic manner, i.e., only measurements of the form $P_m$ and $P_{m+1}$ (addition modulo $5$) can be jointly performed. This inequality can be represented by the graph corresponding to a 5-cycle, with the vertices denoting measurements and edges denoting mutual compatibility; the non-contextual bound of $2$ being given by the independence number of the graph~\cite{Winter} (Independence number is the maximal number of mutually disconnected vertices in the graph). Quantum mechanically, the $P_m$ correspond to projective measurements obeying the cyclic orthogonality condition $P_m P_{m+1} = 0$. 

It was shown in~\cite{Badziag} that for this inequality the spectrum of the KCBS operator $\hat{K} = \sum_m P_m$ can be realized as a function of a single parameter $-1 \leq s \leq 1$,
\begin{align}
& K_{max} = \frac{3 + s^2}{2} + \frac{1}{2}\sqrt{\frac{1 + 3 s^2 - 5 s^4 + s^6}{1 + s^2}}; \label{kmax} \\
& K_2 = \frac{3 + s^2}{2} - \frac{1}{2}\sqrt{\frac{1 + 3 s^2 - 5 s^4 + s^6}{1 + s^2}}; \label{k2} \\
& K_3 = 2 - s^2. \label{k3}
\end{align}
$K_{max}$ is the largest eigenvalue of the operator taking values between $2$ and $\sqrt{5}$. For $0 \leq s^2 \leq \frac{\sqrt{5}-1}{2}$, $2 \geq K_3 \geq K_2 \geq 1$ and in the parameter regime $\frac{\sqrt{5}-1}{2} \leq s^2 \leq 1$, $2 \geq K_2 \geq K_3 \geq 1$. It can be seen that $K_{max} + K_2 + K_3 = 5$. Let us reiterate that the maximal value of the KCBS expression is then obtained as $\max \trace[\hat{K} \rho] = K_{max} \rho_{max} + K_{mid} \rho_{mid} + K_{min} \rho_{min}$ where the spectrum of the KCBS operator is given by $K_{max} \geq K_{mid} \geq K_{min}$ and the eigenvalues of the qutrit state are ordered as $\rho_{max} \geq \rho_{mid} \geq \rho_{min}$. In fact, further simplification is possible by recognizing that $K_{min} = 5 - K_{max} - K_{mid}$ and $\rho_{min} = 1 - \rho_{max} - \rho_{min}$. In the next section, we utilize the above parametrization and show a connection between the set of symmetric two-qubit states that violate KCBS inequalities and the set of corresponding states that lead to a violation of the simplest Bell inequality, namely the CHSH inequality~\cite{CHSH}. 

Physically we are constrained to measure observables of the form of~(\ref{3obs}). We may construct the following rank-one projections from such observables: $\frac{1}{2}(S_m \mp S_m^2)$ and $\mathbb{I} - S_m^2$ corresponding to $|S_m=\pm 1 \rangle$ and $|S_m=0\rangle$ respectively. In~\cite{Kurzynskietal} it is shown that, given restriction of SO(3) operations, the projection $\frac{1}{2}(S_m + S_m^2)$ can only be measured in one context, namely 
\begin{gather}
\frac{1}{2}(S_m + S_m^2), \quad \frac{1}{2}(S_m - S_m^2), \quad \mathbb{I} - S_m^2. 
\end{gather}
This is not sufficient to form a KCBS inequality, as we need five rank-one projectors $\{P_m\}$ such that $P_m$ is measured in two different contexts, $P_{m\pm 1}$, such that $P_{m+1}$ and $P_{m-1}$ are not co-measurable.

The original KCBS inequality was proposed for a spin-1 system with projections of the form $P_m=\mathbb{I}-S_m^2$. We may rewrite~(\ref{KCBS1}) as
\begin{gather}\label{KCBS4}
\sum_{m=1}^5  \langle S_m^2 \rangle \geq 3.
\end{gather}
One can find directions $m$ such that for a pure qutrit state, the inequality~(\ref{KCBS1}) is violated up to $\sqrt{5} - 2$, the maximum quantum mechanical value of the left-hand side ($\sqrt{5}$) being the value of the Lovasz-theta function on the 5-cycle graph~\cite{Winter}. In this paper, by KCBS type contextual inequalities, we refer to inequalities of the form~(\ref{KCBS1}) on general graphs. 

\section{Contextual biphotons}

\begin{figure}
\begin{center}
\scalebox{0.3}
{\includegraphics{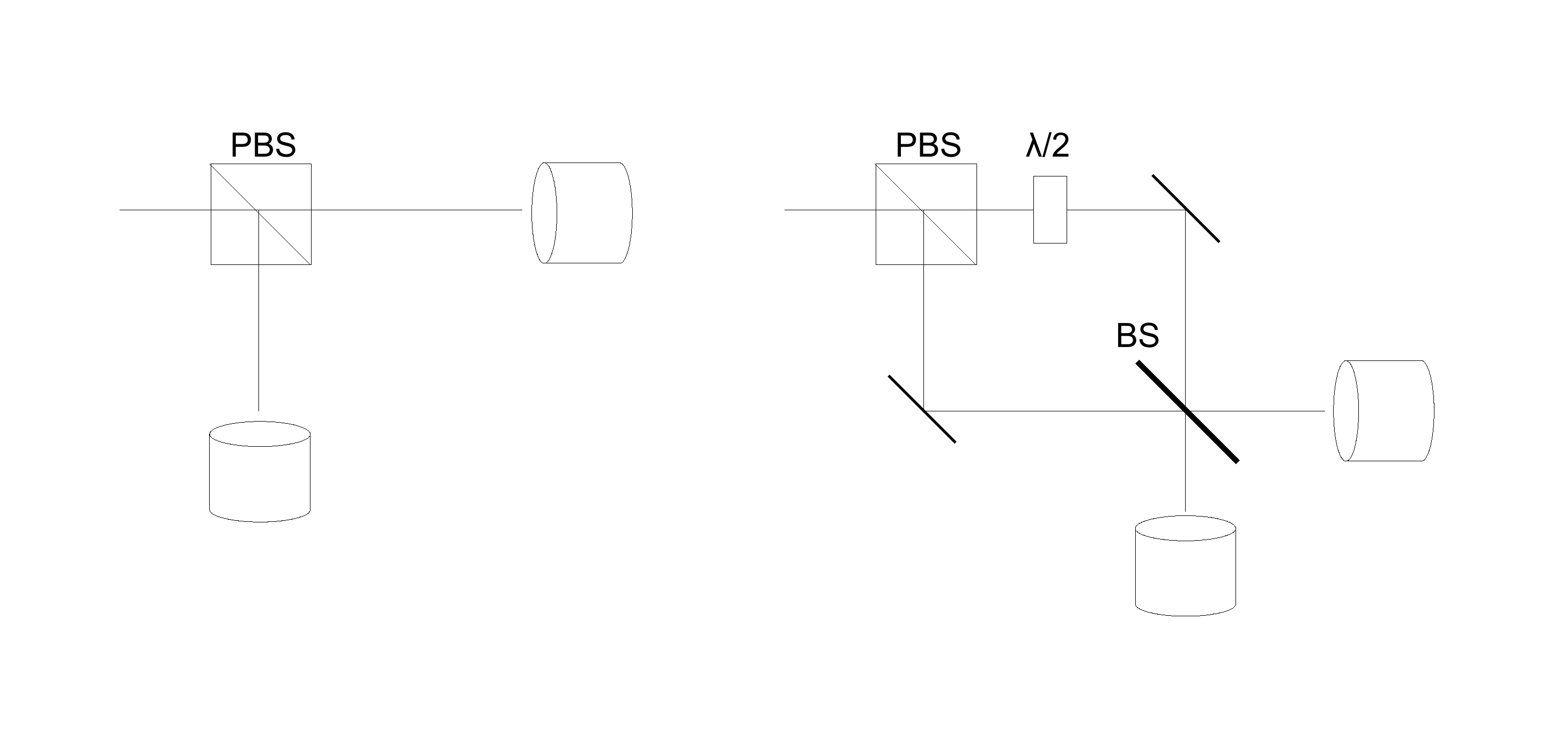}}
\caption{\label{f1a}  An idealized biphoton measurement settings for operators $S_m$ (left) and $S_m^2$ (right).}
\end{center}\end{figure}

The feasible measurement scenarios for biphotons are also limited. One can measure biphoton polarization in a basis  $\{|H,H\rangle,|V,V\rangle,\frac{1}{\sqrt{2}}(|H,V\rangle + |V,H\rangle)\}$ or in any other basis that can be obtained from this one via $SO(3)$ rotations~\cite{SO33}. Note, that these measurements correspond to measurements of spin-1 magnetization $S_m$. In an idealized scenario when one is certain that the state is indeed a state of exactly two photons, such a measurement can be performed using a polarizing beam-splitter (PBS) and two detectors (Fig.~\ref{f1a} left). 
If the upper detector clicks one knows that the biphoton state is $|HH\rangle$, if the lower clicks the state is $|VV\rangle$ and finally if both click the state is $\frac{1}{\sqrt{2}}(|H,V\rangle + |V,H\rangle)$. However, a necessary condition for contextuality is an existence of the measurement context, i.e., an existence of more than one compatible observable. Note, that there are no natural observables that are compatible to $S_m$ and that can be feasibly measured in biphotonic scenario. Fortunately, one can measure a biphotonic observable corresponding to $S_m^2$ that allows for a natural context, since in case of spin-1 operators $[S_m^2,S_n^2]=0$ if $\vec{m}$ is orthogonal to $\vec{n}$. As a result, one can apply~(\ref{KCBS4}) to biphotons.

The measurement of $S_m^2$ in biphotonic scenario can be realized in the following way (Fig.~\ref{f1a} right). The setup resembles the one for the measurement of $S_m$, however this time we need to loose the distinguishability between the two events. The measurement setup resembles the one of an interferometer. After the PBS we put a half-wave plate ($\lambda/2$) in one arm of the interferometer in order to bring photons traveling in both arms to the same polarization state. Next, we perform standard two-photon interference experiment. A simple analysis of the action of the above setup on three orthogonal states
\begin{align}
& \frac{1}{\sqrt{2}}(|H,V\rangle + |V,H\rangle),\\ 
& \frac{1}{\sqrt{2}}(|H,H\rangle + |V,V\rangle),\\ 
& \frac{1}{\sqrt{2}}(|H,H\rangle - |V,V\rangle),
\end{align}
corresponding to $\{|S_z=0\rangle,|S_y=0\rangle,|S_x=0\rangle\}$, respectively, shows that only for the state $\frac{1}{\sqrt{2}}(|H,H\rangle + |V,V\rangle)$ both detectors click with probability $1$, whereas for the two other states one observes that either the upper detector, or the lower detector clicks (in both cases with probability $1/2$). This makes these two states indistinguishable and corresponds to the measurement of $S_y^2$. The measurement of $S_m^2$ for an arbitrary direction $\vec{m}$ can be done using the same setup if a proper $SO(3)$ rotation is performed before the final measurement. 

The inequality~(\ref{KCBS4}) can be also used to establish connection between KCBS and CHSH tests. Since spin-1 state can be realized by a symmetric state of two spin-1/2 particles (biphoton is made of two photons), we can express the spin operators in~(\ref{KCBS4}) using the representation given by~(\ref{square}). After simple manipulations we obtain
\begin{gather}\label{expr}
\sum_{m=1}^5 \langle \vec{m}\cdot\vec{\sigma}\otimes \vec{m}\cdot\vec{\sigma} \rangle \geq 1.
\end{gather}
Note that CHSH expression can be written as~(\ref{BCHSH}) and that all local realistic theories obey $|\langle B_{m,n}\rangle|\leq 2$. It follows that the KCBS inequality can be rewritten as a sum of five CHSH expressions
\begin{gather}\label{iq1}
\sum_{m=1}^5 \langle B_{m,m+1} \rangle \geq 2\sqrt{2}.
\end{gather}

Let us rewrite the KCBS expression~(\ref{expr}) in the following form
\begin{gather}
\sum_{i=1}^3 \lambda_i \left( \sum_{j=1}^5  (\vec{n_i}\cdot\vec{m_j})^2 \right) \geq 1,
\end{gather}
where $\vec{n_i}$ denotes the eigenvector of the correlation tensor corresponding to the eigenvalue $\lambda_i$ and $\vec{m_j}$ ($j=1,\dots,5$) denotes the five directions along which the spin-1 operators for KCBS inequality are defined. This in turn can be written as
\begin{gather}\label{iq2}
\sum_{i=1}^3 \lambda_j (\vec{n_i}^\intercal \cdot \hat{K} \cdot \vec{n_i})  = \trace[\hat{T} \hat{K}] \geq 1,
\end{gather}
where $\hat{T}$ is the correlation tensor and $\hat{K}$ is the KCBS operator. 

Our problem is defined in the following way: {\it Given a quantum state corresponding to $\hat{T}$ is there any $\hat{K}$ that leads to violation of~(\ref{iq2})?} The inequality~(\ref{iq2}) yields
\begin{gather} \trace\left[ 
\begin{pmatrix} \lambda_1 & 0 & 0 \cr 0 & \lambda_2 & 0 \cr 0 & 0 & \lambda_3 \end{pmatrix} 
O \begin{pmatrix} K_{min} & 0 & 0 \cr 0 & K_{mid} & 0 \cr 0 & 0 & K_{max} \end{pmatrix} O^\intercal \right] \geq 1.
\end{gather}
In the above $K_{max}\geq K_{mid} \geq K_{min}$ are eigenvalues of $\hat{K}$ given by~(\ref{kmax}),~(\ref{k2}), and~(\ref{k3}), whereas $O$ is an orthogonal operator representing an $SO(3)$ rotation. We also assume the following ordering of eigenvalues of $\hat{T}$ --- $\lambda_1 \geq \lambda_2 \geq \lambda_3$. Note, that $\lambda_3=1-\lambda_1-\lambda_2$. We have
\begin{gather}
\hat{K}=O \begin{pmatrix} K_{min} & 0 & 0 \cr 0 & K_{mid} & 0 \cr 0 & 0 & K_{max}\end{pmatrix} O^\intercal = \begin{pmatrix} \tilde{k_1} & \star & \star \cr \star & \tilde{k_2} & \star \cr \star & \star & \tilde{k_3} \end{pmatrix}.
\end{gather}
In general, the off-diagonal terms of the above matrix are nonzero. Once again, we can always choose $\tilde{k_3}\geq \tilde{k_1},\tilde{k_2}$. Moreover, $\tilde{k_1}+\tilde{k_2}+\tilde{k_3}=K_{max}+K_{mid}+K_{min}=5$, however due to the interlacing inequality~\cite{HJ} $K_{max}\geq \tilde{k_3}$ and $\min\{\tilde{k_1},\tilde{k_2}\}\geq K_{min}\geq 1$. 

Let us consider the case $\lambda_1 + \lambda_2 \leq 1$ where $1-\lambda_1 -\lambda_2$ is positive. We have
\begin{gather}\label{cond}
\lambda_1 \tilde{k_1} + \lambda_2 \tilde{k_2} + (1-\lambda_1 -\lambda_2) \tilde{k_3} \geq \min\{\tilde{k_1},\tilde{k_2},\tilde{k_3}\}\geq 1,
\end{gather}
therefore it is clear that in this case~(\ref{iq2}) can never be violated. We see that the necessary condition for the violation of~(\ref{iq2}) is the negativity of $\lambda_3$, or in other words, positivity of $\lambda_3$ is a sufficient condition for non-contextuality. Another consequence of this result is that the left hand side of~(\ref{iq2}) is minimal if $\hat{K}$ and $\hat{T}$ share the same eigenbasis. Therefore, the criterion for non-contextuality is
\begin{gather}\label{o}
\lambda_1 K_{min} + \lambda_2 K_{mid} + (1-\lambda_1 -\lambda_2) K_{max} \geq 1.
\end{gather}
Let us now prove a sufficient condition for non-contextuality that will prove particularly interesting from the point of the relation between contextuality and the CHSH inequality --- All states for which $\lambda_1^2 + \lambda_2^2 \leq 1$ do not violate the KCBS inequality~(\ref{iq2}). The proof is a straightforward consequence of the Cauchy-Schwarz inequality. First, let us rewrite~(\ref{o}) as
\begin{gather}
\frac{K_{max} - K_{min}}{K_{max} - 1} \lambda_1 + \frac{K_{max} - K_{mid}}{K_{max} - 1} \leq 1
\end{gather}
Applying the Cauchy-Schwarz inequality to the expression on the left of the above inequality, we obtain that 
\begin{gather}
\frac{(K_{max} - K_{mid})^2 + (K_{max} - K_{min})^2}{(K_{max} - 1)^2}(\lambda_1^2 + \lambda_2^2) \leq 1,
\end{gather}
is a sufficient condition for non-contextuality. The first expression is unambiguously defined with no need to specify the exact expression for $K_{mid}$ and $K_{min}$ in terms of the parameter $s$. A simple analysis in terms of the parameter $s$ shows that this expression has a maximum value of $1$. Therefore, $\lambda_1^2 + \lambda_2^2 \leq 1$ is a sufficient, but as we shall see, not necessary condition for non-contextuality. 

We have shown above that if the sum of squares of the largest eigenvalues of the correlation tensor is bounded by one
\begin{gather}\label{criterion}
\lambda_1^2+\lambda_2^2 \leq 1,
\end{gather}
\begin{figure}
\begin{center}
\scalebox{0.9}
{\includegraphics{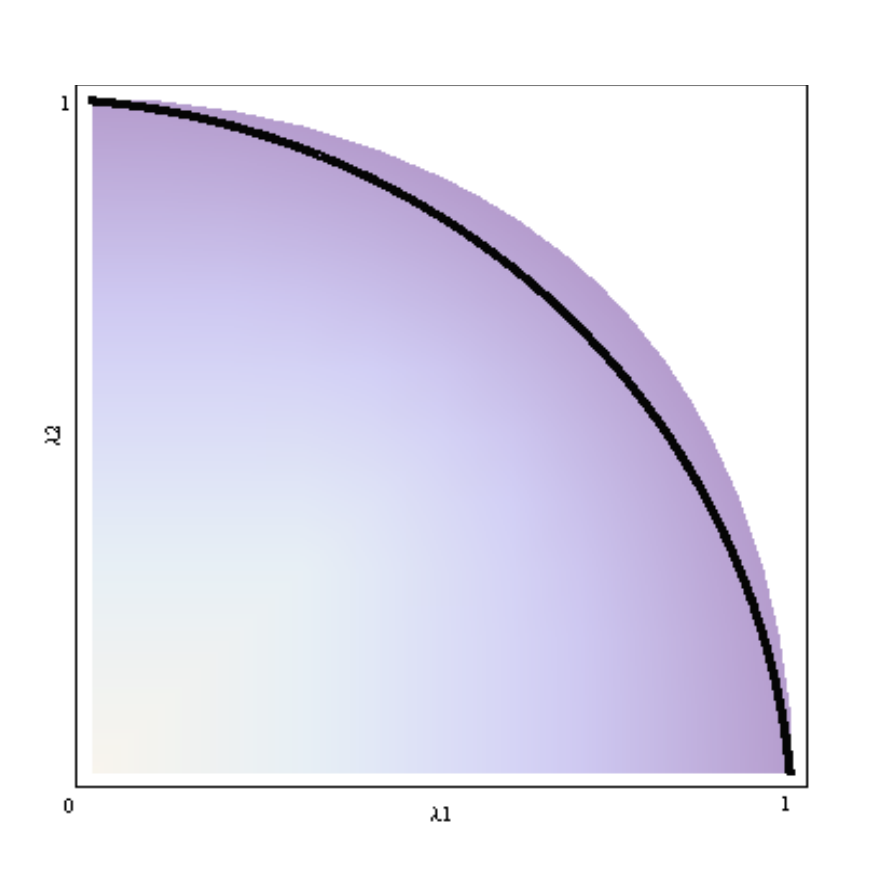}}
\caption{\label{fig1} (Color online) The non-contextual set of states given as a function of $\lambda_1$ and $\lambda_2$. The locality boundary (\ref{criterion}) is represented by the thick black curve.}
\end{center}
\end{figure}
then the state would not exhibit contextuality with respect to~(\ref{iq2}). Note, that the Horodecki criterion~\cite{Hor} states that~(\ref{criterion}) is both necessary and sufficient for a two-qubit state to satisfy any CHSH inequality. Therefore, in our biphotonic scenario locality implies non-contextuality (or contextuality implies nonlocality), however the converse is not true. There exist non-contextual states, with respect to~(\ref{iq2}), which nevertheless do not obey~(\ref{criterion}) and hence are nonlocal with respect to the Horodecki criterion. In this case one can find a suitable CHSH inequality whose respective local measurements lie in the plane spanned by the eigenvectors of $\hat{T}$ corresponding to $\lambda_1$ and $\lambda_2$. Note, that since $\hat{K}$ and $\hat{T}$ share the same eigenbasis and due to minimization orderings~(\ref{o}), the CHSH plane is perpendicular to the eigenvector of $\hat{K}$ corresponding to the largest eigenvalue $k_3$. Due to this reason we refer to this CHSH inequality as to the {\it dual CHSH inequality}.

The set of KCBS non-contextual and CHSH nonlocal symmetric two-qubit states is bounded by two functions of $\lambda_1$ and $\lambda_2$ (see Fig.~\ref{fig1}). The lower bound was derived above~(\ref{criterion}). The upper bound is given by the following parametrization
\begin{align}
\lambda_1 =& \frac{-1+s^2+s^4-(1+s^2)^2\sqrt{9+(s-6)s-\frac{8}{1+s}}}{-2+2s(-1+2(-1+s)s)}\\
\lambda_2 =& \frac{(-2+s)s(1+s)}{-2+2s(-1+2(-1+s)s)}.
\end{align}
As can be seen in Fig.~\ref{fig1}, the upper bound is represent by the curve which is more convex than the circle and the region of non-local and non-contextual states has the crescent moon shape. The width of this region can be estimated by finding the boundary point corresponding to $\lambda_1=\lambda_2$. For the local/nonlocal boundary $\lambda_1=\lambda_2=\frac{1}{\sqrt{2}}\approx 0.707$, whereas for contextual/non-contextual boundary we find that $\lambda_1=\lambda_2=\frac{5+\sqrt{5}}{10}\approx 0.724$.

\section{Discussion}

Under the restriction to the same $SO(3)$ operations on two photons, it was found that locality in the CHSH scenario implies non-contextuality in the KCBS scenario. If the restriction is lifted and general $SU(3)$ transformations are allowed on the total biphotonic system, as is well-known one can find operators that lead to a violation of the KCBS inequality for any pure state and therefore even separable states of two photons can lead to contextuality. The restricted KCBS operators considered here are motivated by the fact that in many implementations precise control over operations of two qubits is more difficult than single-qubit operations. In biphoton systems we firstly lose addressability due to the indistinguishability of the particles and secondly, the general $SU(3)$ operations require interactions between the two photons that are hard to engineer and control. 

It would be interesting to investigate if the general relationship between locality and non-contextuality found here extends to the general situation involving more qubits. For instance, does locality with respect to a two-setting three-qubit Bell inequality such as the Mermin inequality~\cite{Mermin} imply non-contextuality for a state-dependent contextual inequality for a spin-3/2 system? Nonlocality is a well-established resource leading to quantum advantage in tasks as diverse as cryptography, randomness amplification, etc.~\cite{Ekert, Acin}, it is therefore interesting to check if in these or other applications, general single system contextuality is a meaningful resource.   

{\bf Acknowledgments.}  We acknowledge stimulating discussions with G. A. Maslennikov on biphotons. This research is supported by the National Research Foundation and Ministry of Education in Singapore. P. K. acknowledges financial support from the Foundation for Polish Science. K.M. is supported by the John Templeton Foundation.

\end{document}